\documentclass{article}
\usepackage{booktabs}
\usepackage{pdflscape}
\usepackage{hhline}
\usepackage{amsmath,amssymb}
\usepackage{mathtools}
\usepackage{multirow}
\usepackage[most]{tcolorbox}
\usepackage[table]{xcolor}
\usepackage{xcolor}

\newcommand{\unitywspace}{\quad 1\,}

\newcommand{\myhigh}{\cellcolor{gray!30}}

\usepackage{hyperref}

\title{The new SI and its fundamental constants}
\title{Organizing the new SI}
\title{Organizing units according to the new SI} 
\title{The structure of the new SI}
\author{{ B.C.~Regan}\\
	{\small \emph{Department of Physics \& Astronomy and California NanoSystems Institute,}} \\ {\small \emph{University of California, Los Angeles, California 90095 USA}}}

\date{}

\begin{document}
	\maketitle
	
	\begin{abstract}
The ``new''\emph{Syst\`eme international d'unit\'es} (SI), which became effective May 20, 2019, defines and is defined by a set of constants. These include the speed of light, the Planck constant, the Boltzmann constant, and the constant relating the elementary electric charge to the coulomb.  Interpreting such constants as conversion factors organizes the units they relate into a unifying geometric framework. In this framework, units appear (perhaps raised to some power) either as rows/columns in a single conversion table or as entries in a list of dimensionless numbers. This organization clarifies the distinction between ``fundamental'' physical constants with values that are set by people, like those defined in the SI, and those with values that are set by nature.  It also reveals geometry permeating our theories of physics that is normally hidden by a surplus of units.
	\end{abstract}

	 The \emph{Syst\`eme international d'unit\'es}, or ``SI'' for short, is  the standard units system for commerce and science worldwide.  This system is the direct descendant of the decimal system of weights and measures set up in France in the 1790's during the times of the French Revolution. In its original form, this system defined two base units ---  the meter and the gram --- in terms of the Earth's circumference and the density of water.  Over the years the number of base units has been increased to seven and the definitions have been updated to take advantage of scientific progress.
	 
	 Recently the SI has  gone through a radical change.  As of 2019, the core of the SI is no longer  a list of defined units, but rather a list of defined constants.  This formulation of the SI --- the ``new SI'' --- is a major advance, in that it decouples the unit definitions from their realizations, freeing the realizations from any particular technology or implementation.  Most notably, it frees the SI from its last prototype artifact by redefining the kilogram in terms of Planck's constant.
	 
	 The re-formulation of the SI based on this new  logical structure encourages us to re-examine the entire edifice. The Bureau International des Poids et Mesures (BIPM, known in English as the International Bureau of Weights and Measures) publishes a brochure  \cite{inglis_si_2019} to promote and explain the SI. For ease of reference we quote the SI brochure here:
	 \begin{tcolorbox}[breakable, enhanced] \setlength{\parindent}{20pt}
	 {\large\noindent\textbf{Box~1}}\newline
	 	The International System of Units, the SI, is the system of units in which
	 	\begin{itemize}
	 		\item 	The unperturbed ground state hyperfine transition frequency of the caesium
	 		133 atom, $\Delta \nu_\text{Cs}$, is $9\,192\,631\,770$~Hz.
	 		\item the speed of light in vacuum, c, is $299\,792\,458$~m/s,
	 		\item the Planck constant, $h$, is $6.626\,070\,15 \times 10^{-34}~\text{J s}$,
	 		\item the elementary charge, $e$, is $1.602\,176\,634 \times 10^{-19}~\text{C}$,
	 		\item the Boltzmann constant, $k$, is $1.380\,649 \times 10^{-23}~\text{J/K}$,
	 		\item the Avogadro constant, $N_A$, is $6.022\,140\,76 \times 10^{23}~\text{mol}^{-1}$,
	 		\item the luminous efficacy of monochromatic radiation of frequency $540 \times 10^{12}$~Hz, $K_\text{cd}$, is 683~ lm/W,
	 	\end{itemize}
	 	where the hertz, joule, coulomb, lumen, and watt, with unit symbols Hz, J, C, lm, and W,
	 	respectively, are related to the units second, metre, kilogram, ampere,  kelvin, mole, and candela,
	 	with unit symbols s, m, kg, A, K, mol, and cd, respectively, according to Hz = s$^{–1}$,
	 	J = kg m$^{2}$ s$^{-2}$, C = A s, lm =cd sr,  and W = kg m$^2$ s$^{-3}$.
	 \end{tcolorbox}

While the above list of defining constants completely specifies the core of the SI,  it implies an encompassing logical structure that this list-based format does not make immediately obvious.   To give just one example, it is not obvious how this list of constants defines the kilogram.  It is the purpose of this communication to show that these constants, as well as the electric constant $\epsilon_0$  and the magnetic constant $\mu_0$ that were defined in previous versions of the SI, can be thought of as conversion factors. 	The SI has always been a system of units, but it is now explicitly also a system for converting \emph{between} units. As such, it can be organized in a square table or matrix. Where  the list-based format gives  five, seemingly isolated  pairwise relationships between a handful of SI units, a table-based format connects and organizes almost all of them. 

 Throughout the course of human history, measurement systems like the SI have arisen in response to the needs of the moment.  Frequently it has been the case that different units are used to measure quantities of the same type in different  contexts. The cubit, the foot, the meter, the fathom, and the mile --- these units of distance (and many more) were each defined by a person or group of people according to some perceived convenience.  These conveniences have varied  from instance to instance and from culture to culture, and the units themselves vary correspondingly.
	
	 With its long history of quantification, distance (or length) provides an excellent example of how a multitude of  units have arisen to describe similar physical quantities. Here `similar' might mean identical --- for instance, horizontal distance might be given in kilometers or miles. Or it might not --- altitude and depth have distinct meanings as distance above and below some surface, respectively, and might be given in feet or fathoms, respectively. This concept of similarity will prove increasingly important as we move forward.  Altitude and depth are not the same, and we can highlight this difference by giving them in different units.  But they are similar, and, if we choose, we can give them both in the same units --- say, meters --- without any essential loss of information.
	 
	 \begin{table}
	 	\centering
	 	\resizebox{\linewidth}{!}{
	 		\begin{tabular}{|l|c|c|c|c|c|c|c|c|}
	 			\toprule
	 			& cm & inch & feet & cubit & meter & fathom & li & mile \\
	 			\midrule
	 			cm & \textbf{1} & 0.394 & 0.033 & 0.019 & 0.010 & 5.47E-03 & 2.00E-05 & 6.21E-06 \\
	 			inch & \textbf{2.54} & \textbf{1} & 0.083 & 0.049 & 0.025 & 0.014 & 5.08E-05 & 1.58E-05 \\
	 			foot & 30.480 & \textbf{12} & \textbf{1} & 0.588 & 0.305 & 0.167 & 6.10E-04 & 1.89E-04 \\
	 			cubit & \textbf{51.86} & 20.417 & 1.701 & \textbf{1} & 0.519 & 0.284 & 1.04E-03 & 3.22E-04 \\
	 			meter & \textbf{100} & 39.370 & 3.281 & 1.928 & \textbf{1} & 0.547 & 2.00E-03 & 6.21E-04 \\
	 			fathom & 182.880 & 72.000 & \textbf{6} & 3.526 & 1.829 & \textbf{1} & 3.66E-03 & 1.14E-03 \\
	 			li & 5.00E+04 & 1.97E+04 & 1.64E+03 & 964.134 & \textbf{500} & 273.403 & \textbf{1} & 0.311 \\
	 			mile & 1.61E+05 & 6.34E+04 & \textbf{5280} & 3.10E+03 & 1.61E+03 & 880.000 & 3.219 & \textbf{1} \\
	 			\bottomrule
	 		\end{tabular}
	 	}
	 	\caption{\textbf{Units of length.} The table can be read \{table entry\} [column heading] per [row heading], or one [row heading] is \{table entry\} [column heading].  For instance, there are 2.54~cm per inch, and one inch is 2.54~cm. The 7 defining constants are indicated by bold typeface.  With these entries, the rest of the table can be worked out. Numbers that are not bold may be rounded.}
	 	\label{tab:lengths}
	 \end{table}

	Table~\ref{tab:lengths} provides conversion factors between a few different units of length. Such tables,  probably familiar to most readers, are easily constructed. A $2\times 2$ table is fully defined by one conversion constant.  If, say, there are $d=2.54$~cm/in, then 
	there are $1/d= (1/2.54 \simeq 0.3937\ldots)$~\mbox{in/cm}, which accounts for the off-diagonal elements. (Here we have defined the constant ``$d$'' for ease of future reference.) The diagonal elements are unity, because there is 1~\mbox{in/in} and 1~\mbox{cm/cm}.  A conversion factor connecting a new unit to any unit already in the table allows the new unit to be added. For instance, with the additional knowledge that there are 100~cm/m, a $3\times 3$ table connecting the centimeter, the inch and the meter  can be constructed.  Thus $N$ connected conversion factors fully define an $(N+1)\times (N+1)$ unit conversion table.
	
Several observations can be made here that might seem trivial in the present context.  First, none of the constants appearing in Table~\ref{tab:lengths} would be described as ``fundamental''.  Rather, they are the nearly random products of culture and history.  Second, any one of them can be --- and in fact often is --- set equal to unity.  Given a length $L$ in inches, we multiply by $1=d=2.54$~cm/in to convert this length to centimeters.  Setting a constant in the table equal to unity is precisely the mechanism by which we convert units.  Third, one can reasonably consider each of the table entries as either a  dimensioned number, or as a dimensionless one.  For instance, $d$ has dimensions of cm/in in one sense, and no dimensions as a length per length in another. Fourth, grouping different units together in a conversion table represents a unification of sorts, one that allows us to rank, convert, and compare units that we might otherwise find in different, perhaps ``dimensionally incompatible'' contexts. The table shows that, while we might not choose to measure altitude in fathoms for cultural or historical reasons, we could if we wanted to. The existence of these inter-conversion factors implies that all of these units are, putting cultural considerations aside, interchangeable and, in that sense, equivalent. Moreover, distinct  physical quantities described in terms of these units fit within an encompassing logical structure.  Depth and altitude are not the same, but for the price of a minus sign we can unify them within a broader context, position.  We can acknowledge that depth and altitude are not the same while at the same time deciding to measure both in the same unit.
	
As a prelude to addressing the full SI, we now extend Table~\ref{tab:lengths} to accommodate the SI's definition of the meter. Since 1983,  the meter has been effectively defined in terms of a time and a speed: the second and the speed of light, $c$, respectively. The meter is the distance that light travels in vacuum in $1/299 792 458^\text{th}$ of a second.  Inspecting the units of $c$, it is not difficult to see how this quantity connects the meter and the second, which is to say distance and time.  In fact, the constant $c$ follows the same $x$-per-$y$ pattern as the constant $d$ that connects the centimeter and the inch.
	
	Because the $x$-per-$y$ pattern is so clearly that of a conversion factor, there are no mathematical difficulties in constructing a table that connects time and distance units (Table~\ref{tab:MetersSecondsYears}). One might even argue that it is only classical preconceptions that stop us from seeing $c$ as a conversion factor, first and foremost. However, we are introducing some new theoretical assumptions. 	 
	
 Any system of units is laden with theoretical assumptions\cite{wilczek_fundamental_2007,duff_how_2015}. For instance, to define the meter as a general unit of distance, we assume that space is isotropic: it makes sense to measure altitudes, depths, left-right distances, and forward-back distances all in the same units.  Likewise, when we define the second in terms of the $^{133}$Cs hyperfine transition (Box~1), we are assuming that two atomic clocks, perhaps at different altitudes and with some relative motion, can both be correct, even if they disagree\cite{inglis_si_2019}.  This relativistic definition is radically different than a Newtonian definition of the second based on the Earth's rotation, which assumes that one clock (the Earth) suffices for all observers.
 
 When we define the speed of light, we are further committing ourselves to  Einstein's theory of relativity. In the case of $c$, we are assuming that it makes sense to measure time and distance in the same units\cite{1992Taylor}. Space and time are not independent, but part of a unified arena, spacetime.  If we so choose, we can measure distances in time units (e.g.\ the ``light year''), as is common when reporting astronomical distances.  Or we can choose to measure time in distance units (e.g.\ the meter), as is common when reporting the lifetimes of unstable particles\cite{navas_review_2024}.

\begin{table}
	\centering
	\resizebox{\linewidth}{!}{
		\begin{tabular}{|l|c|c|c|c|c||c|c|c|}
			\toprule
			& cm & inch & foot & meter & mile & second & day & year \\
			\midrule
			cm & \textbf{1} & 0.394 & 0.033 & 0.010 & 6.21E-06 & 3.34E-11 & 3.86E-16 & 1.06E-18 \\
			inch & \textbf{2.54} & \textbf{1} & 0.083 & 0.025 & 1.58E-05 & 8.47E-11 & 9.81E-16 & 2.68E-18 \\
			foot & 30.480 & \textbf{12} & \textbf{1} & 0.305 & 1.89E-04 & 1.02E-09 & 1.18E-14 & 3.22E-17 \\
			meter & \textbf{100} & 39.370 & 3.281 & \textbf{1} & 6.21E-04 & 3.34E-09 & 3.86E-14 & 1.06E-16 \\
			mile & 1.61E+05 & 6.34E+04 & \textbf{5280} & 1.61E+03 & \textbf{1} & 5.37E-06 & 6.21E-11 & 1.70E-13 \\
						\hhline{|-|=|=|=|=|=|=|=|=|}
			second & 3.00E+10 & 1.18E+10 & 9.84E+08 & \textbf{299792458} & 1.86E+05 & \textbf{1} & 1.16E-05 & 3.17E-08 \\
			day & 2.59E+15 & 1.02E+15 & 8.50E+13 & 2.59E+13 & 1.61E+10 & \textbf{86400} & \textbf{1} & 2.74E-03 \\
			year & 9.46E+17 & 3.72E+17 & 3.10E+16 & 9.46E+15 & 5.88E+12 & 3.16E+07 & \textbf{365.25} & \textbf{1} \\
			\bottomrule
		\end{tabular}
	}
	\caption{\textbf{Units of length and time.} Like  the other tables in this article except Tables~\ref{tab:dimensionless} and \ref{tab:EHBDsplit}, this table is an extension of Table~\ref{tab:lengths} that is abbreviated for display purposes.  It connects distance units (upper left $5\times 5$) with time units (lower right $3\times 3$). One value off  the block diagonal, $c$, connects the two sub-tables. ``Year'' refers to the Julian year, which is preferred by astronomers. Notable entries include $c$ in imperial units  (186,000 miles $\simeq 1$~second), 1~foot $\simeq 1$~ns, and the useful approximation 1~yr $\simeq\pi\times 10^7~$s.}
	\label{tab:MetersSecondsYears}	
\end{table}

	We can be more explicit about the meaning of ``makes sense'' in the preceding paragraph. In the Newtonian system of time independent of space, one could perfectly well define a conversion factor analogous to the speed of light that would allow, say, distances to be given in time units.  However, there is little motivation  (``it would not make sense'') to do so, because in Newtonian physics there is no way to inter-convert space and time.  What makes defining $c$ sensible in our modern (relativistic) theory of physics is the fact that, while time and distance are not the \emph{same}, they are not wholly distinct either. Space and time are fungible, in much the same way that altitude and depth are fungible with respect to sea level in the presence of tides. They can be mixed via a Lorentz transformation, which is to say that two observers with some relative motion can reasonably disagree on time and space intervals considered separately. 
	
	Taylor and Wheeler\cite{1992Taylor} give a ``Parable of the Surveyors'' that beautifully illustrates this idea with a purely spatial analogy. In their parable, a culture exists that measures east-west distances in meters and north-south distances in miles. An east-west distance is not the same as a north-south distance, but one can be converted into the other, either partially or wholly, by the choice of orientation of the coordinate axes. In the parable,  some Surveyors (`nighttime') orient their N-S coordinate axis towards true north and others (`daytime') orient their N-S coordinate axis towards magnetic north. Surveyors from the two camps can reasonably disagree on north-south and east-west separations considered separately. 
	
	This Surveyor culture exists in region small enough that the Earth's curvature is negligible.  Thus daytime and nighttime surveys can be reconciled via Euclidean rotations (which happen to be particular instances of a Lorentz transformation).  The calculations are complicated by appearance of  conversion factors between meters and miles. It ``makes sense'' to measure east-west and north-south distances in the same units, because  the distinctions are fungible.  Using the same units simplifies the transformation laws and makes the fungibility explicit.   
	
	As Taylor and Wheeler further argue, the same reasoning applies to space and time. Like the Surveyors, we too have cultural hang-ups.  We overwhelmingly consider time and distance  to merit different units, just as the Surveyors consider east-west and north-south distances to merit different units.   Just as a Euclidean $(x,y)$ rotation is simplified if $x$ and $y$ are given in the same units, so a Lorentz $(t,x)$ boost (i.e.\ a Minkowski rotation) is simplified if $t$ and $x$ are given in the same units. By maintaining distinct, ``sacred'' units for space and time, we obscure the underlying geometry, exhibiting the same cultural backwardness as the Surveyors.  
	
	Table~\ref{tab:MetersSecondsYears} thus gives us the freedom to choose. We can stick with tradition, with its benefits of familiarity and widespread usage, and continue to measure time and distance in different units.  Or we can, in cases where the circumstances so warrant, measure time and distance in the same units. This second approach is mathematically simpler and better adapted to our modern (relativistic) understanding of physics.  Particle physicists and astronomers are both comfortable with the modern approach, for  phenomena  in these fields frequently inter-convert times and distances.  We have no trouble measuring lifetimes in meters or distance in years (although we might refer to the latter unit as a ``light-year'' to prevent any possibility of confusion).  More generally, we are free to choose any number of units from Table~\ref{tab:MetersSecondsYears} to measure all manner of times and  distances. 
	
We are curious to  explore how far the argument of Taylor and Wheeler  might be extended.  Having identified one conversion factor in the SI (Box~1) encourages us to look for more.   The Boltzmann constant, $k$, is obviously of the right form, since its J/K matches the $x$-per-$y$ pattern that we have remarked upon previously. Three others are less obvious.
	
	First, the coulomb's definition does not follow the patterns of $c$ and $k$: it is not given as a conversion factor, nor in terms of a named constant.  But according to this definition $e =1.602\,176\,634 \times 10^{-19}~\text{C}$. Dividing both sides by $e$ gives unity on the left and a familiar conversion factor on the right. Similar reasoning applies to $\Delta \nu_\text{Cs} = 9\,192\,631\,770$~Hz. Finally, we can re-write the J$\cdot$s appearing in the definition of the Planck constant, $h$, as J/Hz \cite{mohr_codata_2025}, which casts it into the form of a conversion factor  --- one that we can relate to the second and to the other units of Tables~\ref{tab:lengths}--\ref{tab:MetersSecondsYears}.

\begin{table}
	\centering
	\resizebox{\linewidth}{!}{
		\begin{tabular}{|l|c|c|c|c|c||c||c|c|}
		\toprule
		& $(\text{kilogram})^{-1}$ & $(\text{kg m/s})^{-1}$ & $(\text{joule})^{-1}$ & $(\text{e volt})^{-1}$ & $(\text{kelvin})^{-1}$ & Cs cycle & meter & second \\
		\midrule
		$(\text{kilogram})^{-1}$ & \textbf{1} & 3.34E-09 & 1.11E-17 & 1.78E-36 & 1.54E-40 & 6.78E-41 & 2.21E-42 & 7.37E-51 \\
		$(\text{kg m/s})^{-1}$ & \textbf{3.00E+08} & \textbf{1} & 3.34E-09 & 5.34E-28 & 4.61E-32 & 2.03E-32 & \textbf{6.63E-34} & 2.21E-42 \\
		$(\text{joule})^{-1}$ & 8.99E+16 & \textbf{3.00E+08} & \textbf{1} & \textbf{1.60E-19} & \textbf{1.38E-23} & 6.09E-24 & 1.99E-25 & \textbf{6.63E-34} \\
		$(\text{e volt})^{-1}$ & 5.61E+35 & 1.87E+27 & 6.24E+18 & \textbf{1} & 8.62E-05 & 3.80E-05 & 1.24E-06 & 4.14E-15 \\
		$(\text{kelvin})^{-1}$ & 6.51E+39 & 2.17E+31 & 7.24E+22 & 1.16E+04 & \textbf{1} & 0.441 & 0.014 & 4.80E-11 \\
		\hline\hline
		Cs cycle & 1.48E+40 & 4.92E+31 & 1.64E+23 & 2.63E+04 & 2.267 & \textbf{1} & 0.033 & 1.09E-10 \\
		\hline\hline
		meter & 4.52E+41 & 1.51E+33 & 5.03E+24 & 8.07E+05 & 69.503 & 30.663 & \textbf{1} & 3.34E-09 \\
		second & 1.36E+50 & 4.52E+41 & 1.51E+33 & 2.42E+14 & 2.08E+10 & \textbf{9.19E+09} & \textbf{3.00E+08} & \textbf{1} \\
		\bottomrule
	\end{tabular}
	}
	\caption{\textbf{Core units of the SI.} This table ranks and relates 8 dimensioned quantities, tying them all to the cesium-133 ground state hyperfine transition. Six are the SI units for mass (kg), momentum, energy and torque,  temperature (K), distance (m), and time (s).  Momentum does not have a named unit in the SI, but it has dimensions (kg$\cdot$m/s) distinct from the other units. Energy appears twice, both as a coherent SI unit (J=kg$\cdot$m$^2$/s$^2$= C$\cdot$V) and as a unit ``accepted for use with the SI'' (eV)\cite{inglis_si_2019}.  Torque has the same dimensions as energy (J), but it is most commonly given in N$\cdot$m. A `per cycle' is understood to accompany all of the inverted headers, e.g.\ kg/cycle. While  rounded here for display purposes,  the numbers in this table are all exact, and are completely determined by the decisions of the BIPM (Box~1). None of them depend on experiment, which is to say that none of them are set by nature.}
\label{tab:newSI}
\end{table}

Using these conversion constants and following the same procedure as before, we extend Tables~\ref{tab:lengths}--\ref{tab:MetersSecondsYears}.  To keep the table a manageable size, we display only the meter  and the second from the first two tables.  The resultant Table~\ref{tab:newSI} provides another perspective on much of the information in Box~1.

Arranged by size, the units  sort themselves into two categories: lengths and frequencies.  (For the remainder of this article we will generally use ``length'' to refer to both distances and times.  Likewise ``frequency'' will refer to both temporal and spatial frequencies.)  On the left are the high frequencies and small lengths, and on the right the low frequencies and large lengths.  All of the reciprocal (i.e.\ frequency) space quantities are on the left; a kilogram is bigger than a joule is bigger than an electronvolt is bigger than a kelvin. The coherent SI unit of momentum lies between the kilogram and the joule.  

The upper left block consists of quantities native to reciprocal space, e.g.\ energy and momentum, while the lower right block consists of real space quantities, e.g.\ time and distance.  It is not entirely coincidental that the Cs cycle happens to be numerically on the boundary between them. Our species is crossing a technological threshold from  low (i.e.\ classical) frequencies toward  other ``everyday'' scales such as the joule and the kilogram just as we realize (with the development of quantum mechanics) that these scales can be interpreted as frequencies.

As a phase, the Cs cycle most obviously takes equally from both spaces.  In other words, the $^{133}$Cs ground state hyperfine transition, chosen because of its suitability for atomic clocks,  can be thought of as defining a time and/or a frequency.  Of course, every other quantity in the table also defines a time and a frequency, but the cesium transition is the only direct reference to a natural scale in the table. The other quantities previously referenced natural scales (e.g.\ the Earth's size), but those connections are now severed. 


The bold quantities in Table~\ref{tab:newSI} define the table and are taken from the SI Brochure\cite{inglis_si_2019} (Box~1). The International Bureau of Weights and Measures (BIPM) and other standards organizations such as the National Institute of Standards and Technology (NIST) and the Committee on Data of the International Science Council (CODATA) refer to the Planck constant $h$ and the speed of light in vacuum $c$, which appear in the table, as ``fundamental physical constants''\cite{inglis_si_2019,mohr_codata_2025}. While this language is almost universally adopted, constants like $c$ and $h$ are fundamental physical constants in the same way that   $f=12$~in/ft in Table~\ref{tab:lengths} is a fundamental physical constant.  They relate physical quantities (e.g.\ lengths), and they are fundamental to their  unit systems (SI and imperial, respectively).  However,  while  $h$ and $c$ are ``fundamental'' in this narrow sense, they are  not universal, natural, or unchanging. No alien scientist visiting from another planet would recognize these numbers.  None of the defining constants (Box~1) of the SI have any significance outside of the SI. The BIPM can change, and in fact has changed, their values.  Where it employs the term ``defining constant'' in lieu of ``fundamental constant'', the Brochure is being clearer.

Another way to understand that the numbers in Box~1 are not fundamental aspects of nature is to note that they are not necessary components of our description\cite{regan2020ahistoricalapproachelementaryphysics}.  If we  measure time in meters, we no longer need a constant $c$.  If we also  measure mass and energy in inverse meters, we no longer need a constant $h$.  Meters need not be the preferred unit, for any of the other units, or sets of those units, would do. Particle physicists, for instance,  measure mass in GeV (an energy unit) and lifetimes in meters (a distance unit) as noted above. Of the SI units, we prefer the meter as a general-purpose unit, because powers of length have geometric interpretations and  associated vocabulary (e.g. m$^2$ is an area) that are useful and not available with other units.

\begin{table}
	\centering
	\resizebox{\linewidth}{!}{
		\begin{tabular}{|l|c|c|c|c|c|c|c|c|}
			\toprule
			& $(\text{joule})^{-1}$ & $(\text{newton})^{-1/2}$ & ($\text{watt})^{-1/2}$ & $(\text{pascal})^{-1/4}$ & $(\text{liter})^{1/3}$ & meter & $(\text{hectare})^{1/2}$ & second \\
			\midrule
			$(\text{joule})^{-1}$ & \textbf{1} & 4.46E-13 & 2.57E-17 & 2.98E-19 & 1.99E-24 & 1.99E-25 & 1.99E-27 & \textbf{6.63E-34} \\
			$(\text{newton})^{-1/2}$ & 2.24E+12 & \textbf{1} & 5.78E-05 & 6.68E-07 & 4.46E-12 & 4.46E-13 & 4.46E-15 & 1.49E-21 \\
			($\text{watt})^{-1/2}$ & \textbf{3.88E+16} & 1.73E+04 & \textbf{1} & 0.012 & 7.72E-08 & 7.72E-09 & 7.72E-11 & 2.57E-17 \\
			$(\text{pascal})^{-1/4}$ & 3.36E+18 & 1.50E+06 & 86.511 & \textbf{1} & 6.68E-06 & 6.68E-07 & 6.68E-09 & \textbf{2.23E-15} \\
			$(\text{liter})^{1/3}$ & 5.03E+23 & 2.24E+11 & 1.30E+07 & 1.50E+05 & \textbf{1} & 0.100 & 1.00E-03 & 3.34E-10 \\
			meter & 5.03E+24 & 2.24E+12 & 1.30E+08 & 1.50E+06 & \textbf{10.00} & \textbf{1} & \textbf{0.01} & 3.34E-09 \\
			$(\text{hectare})^{1/2}$ & 5.03E+26 & 2.24E+14 & 1.30E+10 & 1.50E+08 & 1000.000 & \textbf{100.00} & \textbf{1} & 3.34E-07 \\
			second & 1.51E+33 & \textbf{6.73E+20} & \textbf{3.88E+16} & 4.49E+14 & 3.00E+09 & \textbf{3.00E+08} & 3.00E+06 & \textbf{1} \\
			\bottomrule
		\end{tabular}
	}
	\caption{\textbf{Derived units and units-accepted-for-use with the SI.} The newton, watt, pascal, liter, and hectare are proportional to powers  not $\pm 1$ of the kilogram,  meter, and  second. As in Tables~\ref{tab:lengths}--\ref{tab:newSI}, all of these table entries are exact, although here again many are rounded for display purposes.  In the same way that  hectares are areas in real space,  newtons (J/m) and watts (J/s) are areas in reciprocal (frequency) space.}
	\label{tab:PowersOfLength}
\end{table}

When, to incorporate  more of the named SI units, we further extend the approach that led to Table~\ref{tab:newSI}, such geometric interpretations arise automatically.  Table~\ref{tab:PowersOfLength} shows how other ``mechanical'' units acquire  geometric meanings.  Ascribing dimension of 1/length (i.e. a frequency) to the joule gives the watt (W=J/s) dimension 1/length$^2$, a reciprocal area.  Similarly, the newton (N=J/m) and the pascal (Pa = N/m$^2$) can be cast into units of 1/length$^2$ and 1/length$^4$, respectively. Because the relationships between the joule, the meter, and the second are exactly defined, the geometric relationship between, say, the watt and the joule (or the watt and the meter) is as exactly known as the geometric relationship between the hectare and the meter.

Not all SI units have such exact interrelationships. In the new SI, the electromagnetic units are related to the units already discussed by the fine-structure constant, $\alpha  = e^2/2\epsilon_0 h c$, a dimensionless number.  This constant's value cannot be  defined any more than the ratio of circle's radius to its circumference can be defined.  However, unlike $2\pi$,  $\alpha$ cannot be calculated with present theory; it must be measured. Given its definition in terms of four named constants, various routes forward present themselves. The BIPM has experimented with more than one of them. Previously, $\epsilon_0$ and $c$ were defined and $h$ and $e$ were uncertain. In the present ``new'' SI, $h$, $c$, and C$/e$ are defined, and $\epsilon_0$  inherits the uncertainty in $\alpha$.

\begin{table}
	\centering
	\resizebox{\linewidth}{!}{
		\begin{tabular}{|l l ll|}
			\toprule
		\multirow[c]{2}{*}{name}	 & 	\multirow[c]{2}{*}{formula} & \multicolumn{2}{c|}{value} \\
			&&	dimensioned& dimensionless\\
			\hline\hline
			fine structure constant & $\alpha = e^2/2\epsilon_0 h c$& \multicolumn{2}{c|}{$ 7.297\, 352\, 5643(11)\times 10^{-3}$}\\
			&$1/\alpha$ &\multicolumn{2}{c|}{$ 137.035\, 999\, 177(21)$}\\
			\hline
			electronic charge &	$\sqrt{2 \alpha\epsilon_0 h  c} = \sqrt{2 \alpha}$ &  $\unitywspace e$&$ 0.120\,808\,5474\ldots$\\
			&$1/e$& $\unitywspace e^{-1}$& $8.277\,560\,002\ldots$ \\
			coulomb& $ e (\text{C}/e)$&$\unitywspace$C&$7.540\,276\,449\ldots \times 10^{17}$\\
			ohm  &$\frac{\text{V}}{\text{A}}= \frac{\text{J}}{\text{C}}\frac{\text{s}}{\text{C}}$ &$\unitywspace\Omega$&$0.002\,654\,418\,729\ldots$\\
			siemens&$ 1/\Omega $ &$\unitywspace$S&$ 376.730\,3134\ldots$\\
			weber & $\text{V s}= \text{J s}/\text{C}$&$\unitywspace$Wb&$2.001\,505\,103\ldots\times 10^{15}$\\
			gray & $\text{J}/\text{kg}=\text{m}^2/\text{s}^2$&$\unitywspace$Gy&$1.112\,650\,056\ldots\times 10^{-17}$\\
			\hline
			conductance quantum &$G_0 = 2e^2/h$ &$7.748\,091\,729\ldots\times 10^{-5}\,$S &0.029\,189\,4102$\ldots$\\		
			von Klitzing constant & $R_K = h/e^2$&$2.581\,280\,745\dots\times 10^{4}\,\Omega$ &68.517\,999\,58$\ldots$\\
			magnetic flux quantum & $\Phi_0 = h/2e $&$2.067\,833\,848\ldots\times 10^{-15}\,$Wb &4.138\,780\,001$\ldots$\\
			Josephson constant & $K_J=2e/h $&$4.835\, 978\, 48 4\ldots\times10^{14}\,$Hz/V &0.241\,617\,0948$\ldots$\\
			Stefan-Boltzmann constant & $\sigma = \frac{2\pi^5}{15}\frac{k^4}{h^3c^2}$&$5.670\,374\,419\dots\times 10^{-8}\,\text{W}/(\text{m}^2\text{K}^4)$ &$40.802\,624\,63\ldots$\\	
			\bottomrule
		\end{tabular}
	}
	\caption{\textbf{SI units and named constants with dimensionless values}. Proportional to meters to the zeroth power, these quantities  cannot appear as rows-slash-columns in the previous tables. Where they relate two units, the dimensionless values themselves  appear as entries in the tables to follow. To arrive at a dimensionless form, we set the conversion constants $c$, $h$, $\epsilon_0$,  $\mu_0$, $Z_0$, and $k$  equal to one.  The gray and the Stefan-Boltzmann constant $\sigma$ are exact. The constants $K_J$, $R_K$, $\Phi_0$, and $G_0$ are exact in their dimensioned form, but not exact in their dimensionless form.   The uncertainty in the inexact numbers derives solely from the experimental determination of $\alpha$, which has a   relative standard uncertainty of  $1.6\times 10^{-10}$\cite{mohr_codata_2025}. }
	\label{tab:dimensionless}
\end{table}

In a third approach, we can set the conversion factors $\epsilon_0$, $h$, and $c$ all equal to unity, which amounts to deciding to measure capacitance, time, length, and inverse energy all in the same units. Then the dimensionless electronic charge $e = \sqrt{2 \alpha} \simeq 0.1208\ldots$ with  uncertainty inherited from $\alpha$.  Since  the coulomb (C) is proportional to the electronic charge $e$, with the conversion factor  defined as exactly $ 1.602\,176\,634 \times 10^{-19}$~C/$e$, it too has a dimensionless value $\text{C}\simeq 7.540\ldots \times 10^{17}$  with  uncertainty inherited from $\alpha$.  

The coulomb and the electronic charge are thus specified units in the new SI, but as dimensionless quantities they do not fit with the other units in Tables~\ref{tab:lengths}--\ref{tab:PowersOfLength}. They are proportional to meters to the zeroth power, and are therefore independent of the meter and  the other units in Tables~\ref{tab:lengths}--\ref{tab:PowersOfLength}.  A few other units such as the ohm, the siemens, the weber, and the gray, share this distinction (Table~\ref{tab:dimensionless}).  The weber $\text{Wb}=\text{V}\cdot\text{s} = \text{J}\cdot \text{s}/\text{C} \simeq 2.00\ldots \times 10^{15}$, while the gray and the sievert are equal: $\text{Gy}=\text{Sv}=\text{J}/\text{kg}= \text{(m/s)}^2\simeq 1.11\ldots\times 10^{-17}$. 

 In this group, only  $e$ has any claim to being set by nature.  The other dimensionless numbers are either pure human inventions (Gy), or scaled by numbers that are human inventions.   The coulomb and the weber are large because macroscopic charges and fluxes are large in comparison to the electronic charge and the magnetic flux quantum, respectively. The gray (and the sievert) are  small because biologically significant radiation doses are small in comparison to the rest  energy (mass) of the creature absorbing the dose. Although the scales of these numbers can be justified on physical grounds, their exact values are arbitrary.
    
Taken at face value, a dimensionless constant --- a pure number --- is the simplest kind of physical quantity.  A selection of  dimensionless constants illustrates how such  quantities can be dressed in terms of a multitude of units (Table~\ref{tab:dimensionless}).  The Josephson constant, the von Klitzing constant, the magnetic flux quantum, and the conductance quantum  are named constants that are proportional to powers of $e$. In the SI they all have different units, none of which mention $e$. In the new SI, their dimensioned values are exact. Their dimensionless values are independent of the SI and inherit uncertainty from $\alpha$. 

In contrast, the Stefan-Boltzmann constant can be calculated exactly. On ``dimensional grounds'' one can estimate that a blackbody at any temperature radiates power roughly corresponding to one thermal photon's worth of energy per thermal photon period per thermal photon wavelength squared, since temperature provides the only scale in the problem.  The constant $\sigma$ represents a complete accounting of the factors of order unity that separate the exact result from the estimate.  Here the temperature scale is also simultaneously an energy scale, a time scale, and a length scale, with  each of these four quantities customarily given in terms of its own unit. If they are all given in terms of the same unit, $\sigma = 2\pi^5/15$ exactly.

Thus  we have at least three distinct types of ``dimensionless'' constants. Some  constants (e.g.\ $d=$ 2.54 cm/in, $c = 299792458$~m/s) are conversion factors with values that are set by convention and that can be set equal to unity by choice of units. Some conversion constants are calculable (e.g.\ $2\pi$ radians/cycle, $\sigma$) and cannot be set equal to unity. Finally, some dimensionless constants cannot be defined or calculated, so we must measure them (e.g.\ $\alpha$, $e$). Numbers in the latter two categories might be recognizable to an alien culture, whereas numbers in the first would not be.

Now that we have developed non-unit powers of length (both zero and non-zero), we extend our tabling to include electromagnetic units (Table~\ref{tab:electromagnetic}), which are legion. In the new SI, electromagnetic units such as the farad, henry, volt, and ampere connect to the non-electromagnetic units such as the meter and the kilogram via $\alpha$, $e$, and C (Table~\ref{tab:dimensionless}).  The farad and the henry are defined by the electric ($\epsilon_0$) and magnetic ($\mu_0$) constants, also known as the permittivity and permeability of free space, respectively. In the previous SI, the values of $\epsilon_0$ and $\mu_0$  were defined such that $\mu_0= 4\pi \times 10^{-7}\,\text{H/m}$ and $\epsilon_0=1/\mu_0 c^2$ and were therefore exact.   However, while the second relation still holds,  in the new SI the first does not. These constants now have some uncertainty that derives from $\alpha$. Inexact though they may be,  $\epsilon_0\simeq 8.854\times 10^{-12}\,\text{F/m}$ and $\mu_0 \simeq 1.257\times 10^{-6}$~H/m still define the farad and the henry in terms of the meter.  The ampere (A = C/s) and the volt (V) can also be related to the meter, the former  via the second and the dimensionless coulomb (C) and the latter via the electron-volt (eV) and the dimensionless electronic charge $e$, respectively. Finally, the volt provides a bridge to electromagnetic fields.  The SI unit for the electric field $\mathbf{E}$ is V/m, which, like the newton and the watt, is an areal frequency.

\begin{table}
	\centering
	\resizebox{\linewidth}{!}{
		\begin{tabular}{|l|c|c|c|c|c|c|c|c|c|}
			\toprule
			& $(\text{joule})^{-1}$ & $(\text{ampere})^{-1}$ & $(\text{volt})^{-1}$ & $(\text{e volt})^{-1}$ & $(\text{V/m})^{-1/2}$ & meter & henry & second & farad \\
			\midrule
			$(\text{joule})^{-1}$ & \textbf{1} & 5.00E-16 & 1.33E-18 & \textbf{1.60E-19} & 5.13E-22 & 1.99E-25 & 2.50E-31 & \textbf{6.63E-34} & 1.76E-36 \\
			$(\text{ampere})^{-1}$ & 2.00E+15 & \textbf{1} & 2.65E-03 & 3.21E-04 & 1.03E-06 & 3.98E-10 & 5.00E-16 & 1.33E-18 & 3.52E-21 \\
			$(\text{volt})^{-1}$ & \textbf{7.54E+17} & 376.730 & \textbf{1} & \textbf{0.12} & 3.87E-04 & 1.50E-07 & 1.88E-13 & 5.00E-16 & 1.33E-18 \\
			$(\text{e volt})^{-1}$ & 6.24E+18 & 3.12E+03 & 8.278 & \textbf{1} & 3.20E-03 & 1.24E-06 & 1.56E-12 & 4.14E-15 & 1.10E-17 \\
			$(\text{V/m})^{-1/2}$ & 1.95E+21 & 9.73E+05 & 2.58E+03 & 312.152 & \textbf{1} & 3.87E-04 & 4.86E-10 & 1.29E-12 & 3.43E-15 \\
			meter & 5.03E+24 & 2.52E+09 & 6.68E+06 & 8.07E+05 & 2.58E+03 & \textbf{1} & \textbf{1.26E-06} & 3.34E-09 & \textbf{8.85E-12} \\
			henry & 4.01E+30 & 2.00E+15 & 5.31E+12 & 6.42E+11 & 2.06E+09 & 7.96E+05 & \textbf{1} & 2.65E-03 & 7.05E-06 \\
			second & 1.51E+33 & \textbf{7.54E+17} & 2.00E+15 & 2.42E+14 & \textbf{7.75E+11} & \textbf{3.00E+08} & 376.730 & \textbf{1} & 2.65E-03 \\
			farad & 5.69E+35 & 2.84E+20 & 7.54E+17 & 9.11E+16 & 2.92E+14 & 1.13E+11 & 1.42E+05 & 376.730 & \textbf{1} \\
			\bottomrule
		\end{tabular}
	}
	\caption{\textbf{Selected electromagnetic SI units}. The ampere and the volt, as  dimensionless quantities times the hertz and the eV, respectively, appear as frequencies, while the henry and the farad are both lengths. The standard unit for electric field $\mathbf{E}$, V/m, is an  inverse area. The impedance of free space, $Z_0= \mu_0 c = 1/\epsilon_0 c = \sqrt{\mu_0/\epsilon_0}\simeq 376.73\ldots\,\Omega$, appears three times here, as an ohm is a volt/ampere, henry/second, and a second/farad. }
	\label{tab:electromagnetic}
\end{table}

 Electromagnetic-field units provide a particularly striking  example of how historical developments have left  an awkward legacy. The  relationships between such quantities as charge, electric potential, and magnetic field  were not immediately obvious, so early measurements were reported in units defined independently. For instance, the potential of a Daniell cell, the resistance of a  column of mercury (or a length of telegraph wire), and the current required to generate a certain force between parallel wires (or a certain silver electrodeposition rate) served as not-necessarily-consistent units for their respective quantities\cite{lynch_history_1985}. Some relationships could be made simple, such as C=A$\cdot$s=V$\cdot$s/$\Omega$=F$\cdot$V and Wb$=$T$\cdot$m$^2$$=$H$\cdot$A$=$H$\cdot$C/s, but not all of them, because the system was over-determined. Later the relationships were understood and  constants such as $\epsilon_0$ and $\mu_0$ were introduced to convert between the by-then-entrenched units.
  
In 1820 \O rsted  deflected a compass needle with an electrical current, demonstrating that electricity and magnetism were related. Maxwell later unified the two subjects into one, showing that an electromagnetic wave in vacuum travels with speed $c=1/\sqrt{\epsilon_0\mu_0}$.  The prescribed relationships between $\epsilon_0$, $\mu_0$, $c$, and $Z_0 = 1/c\epsilon_0$ are the fruit of Maxwell's unification.   The existence of these relationships  indicates that the early unit definitions were more complicated than necessary.
  
 More specifically, the existence of these relationships indicates that we had --- and still have --- more units than necessary. (Updates  to the SI have historically emphasized continuity over simplicity.) The SI unit for $\mathbf{E}$ is V/m, while the SI unit for $\mathbf{D}$ and $\mathbf{P}$ is C/m$^2$. Similarly, for $\mathbf{B}$ the unit is tesla (T), while for $\mathbf{H}$ and $\mathbf{M}$ the unit is A/m. So, for six different fields we have four different units, which are related by the four constants $c$, $\epsilon_0$, $\mu_0$, and $Z_0$. Counting the weber and the meter separately brings the number of different units to six, all related by the web of identities laid out in Table~\ref{tab:EHBDsplit}.  For instance, there are  $377$~volts in an ampere, which is perhaps unsurprising, but there are also $377$~webers in coulomb (Table~\ref{tab:dimensionless}).   

\begin{table}
	\centering
	\resizebox{\linewidth}{!}{
		\begin{tabular}{|l|c|c|c|c|}
			\toprule
			&  $\text{V/m}$ & $\text{A/m}$ & $ \text{T}=\frac{\text{Wb}}{\text{m}^2}$ & $ \text{C/m}^2 $ \\
			\midrule
			$[E] = \text{V/m}$ & \textbf{1} &$1/Z_0$ & $1/c$ & \textbf{8.85E-12} \\
			$[H, M] = \text{A/m}$ & 376.730 & \textbf{1} & \textbf{1.26E-06} & $1/c$ \\
			$[B] = \text{T}=\frac{\text{Wb}}{\text{m}^2}$ & 3.00E+08 &$1/\mu_0$ & \textbf{1} & $1/Z_0$ \\
			$[D, P] = \text{C/m}^2 $ & $1/\epsilon_0$ & 3.00E+08 & 376.730 & \textbf{1} \\
			\bottomrule
		\end{tabular}
	}
	\caption{\textbf{SI units of electromagnetic field}. This table is given in a mixed value/variable format to make its structure more clear. Such a $4\times 4$ conversion table is determined by three conversion constants, but the SI names four: $c$, $\epsilon_0$, $\mu_0$, and $Z_0$. Only two of these are independent.  In the new SI this table is determined by $c$, which is defined directly, and $\epsilon_0$, which is defined in terms of $\alpha$.  The other two constants are set by the relations $\mu_0 = 1/\epsilon_0 c^2$ and $Z_0 = 1/c\epsilon_0$. Thus eight of these relationships are exact and eight have some uncertainty.  As previously, we sort smaller units to the left. Note, however, that these field units all represent areal frequencies. (This identification provides a geometric picture that is in keeping with the conception of field strength as corresponding to the density of field or flux lines.) Therefore in Table~\ref{tab:electromagnetic} A/m, T, and C/m$^2$, or, rather, their inverse square-roots, would be progressively farther to the left of V/m, not farther to the right as shown here. }
	\label{tab:EHBDsplit}
\end{table}

The six electromagnetic fields themselves (as opposed to their units) are related by $\mathbf{D}=\epsilon_0\mathbf{E}+\mathbf{P}$ and $\mathbf{H}=\mathbf{B}/\mu_0 - \mathbf{M}$  in the SI. The constants $\epsilon_0$ and  $\mu_0$  ``fix'' the units here in two different senses: they correct the units to allow for summation, and they cement the particular choice of units. The disadvantages of this approach are clearer in less abstract contexts. For instance, driving from Burlington, Vermont to Montreal, Quebec, one first drives to the US/Canada border ($d_\text{Bb}= 52$~miles) and then from the border to Montreal ($d_\text{bM}=63$~kilometers). The fact that highway distances are customarily given in different units in the two countries does not require us to designate the total distance traveled by $d_\text{BM} = \eta_0 d_\text{Bb} +d_\text{bM}$  ($\eta_0 = 1.609344$~km/mi, Table~\ref{tab:lengths}). Citizens of any country can use $d_\text{BM} =  d_\text{Bb} +d_\text{bM}$. Equality of units is understood as a necessary precondition for addition.

Similarly, we could write $\mathbf{D}=\mathbf{E}+\mathbf{P}$ and $\mathbf{H}=\mathbf{B}- \mathbf{M}$ with two advantages:  the field units could  be chosen freely, and  the field values appearing would be directly comparable. Including an explicit conversion constant such as $\epsilon_0$, $\mu_0$, or $\eta_0$ in these addition laws  fixes the choice of units so as to impede quantitative comparison between the quantities being added. One can finish reading the previous paragraph without learning the distance between Burlington and Montreal, or even which city is farther from the international border. The barriers to a clear mental picture are even higher when the quantities being summed are not distances, but fields, and the  conversion factors are not of order unity, but millionths and trillionths.

Researching ferroelectricity one might, for instance,  encounter a polarization of 10~$\mu$C/cm$^2$ controlled by a coercive field of 1~MV/cm. If one is not simultaneously aware that the former is about 113 times the latter, one misses an opportunity to quantitatively appreciate the technological appeal of ferroelectrics. Fields and areal charge densities are no more different than depths and altitudes. While not the same, they are not so different that they must be measured in different units.

The tables we have considered so far are the products of human decisions. It is worthwhile to consider what sort of table can be constructed from more natural considerations.  Nature provides many scales to choose from beyond the $^{133}$Cs hyperfine transition. The word `scale' here is apt, because `scale' is associated with determining quantity generally, and the word has meanings that are specific to the measurement of mass (or weight), length, distance, volume, and frequency. As the new SI makes explicit, a scale for one of these quantities is a scale for all of them.  Perhaps the most famous natural scales are the masses of  fundamental particles such as the electron and the proton ($m_e$ and $m_p$, respectively).  Other, related scales (e.g.\ the classical electron radius, the Bohr magneton, the nuclear magneton,  and the Rydberg constant) are easily constructed.

An independent scale is provided by Newton's constant of gravitation $G=  6.67430(15)\times 10^{-11}\text{m}^3\, \text{kg}^{-1}\,\text{s}^{-2}$\cite{mohr_codata_2025}. Applying factors of $c$ and $h$ as defined by the SI (Box~1), we can write $G$ in terms of just one SI unit, as opposed to three: $G\simeq 1.64\times 10^{-69}$~m$^2\simeq 3.36\times 10^{14}$~kg$^{-2}\simeq 1.83\times 10^{-86}$~s$^2$. 
Expressing $G$  in terms of one SI unit shows it to be a scale, like $m_e$ and $m_p$, not a conversion factor.  Setting $G=1$ is thus not like setting  one of the defining constants of the SI equal to~1. Within the SI these constants, e.g.\ $c$ and $h$,  cannot be expressed in terms of one SI unit, and setting any one of them equal to one merely declares that some physical quantities are being reported in non-traditional units. We can set a defining constant equal to 1 without leaving the SI, because we can continue to use SI units or a subset of the SI units. 

In contrast, with $c$ and $h$ (or $\hbar$) equal 1,  setting $G=1$ fixes the reference scale, putting $G$ in the role that is occupied by the $^{133}$Cs transition in the SI. The resulting ``Planck'' unit system is independent of the SI and all of its units. Reading from Table~\ref{tab:NaturalConstants}, we see the historical Planck mass, temperature, length, and time in kilograms, kelvin, meters, and seconds, respectively. (A modern convention substitutes $\hbar$ for $h$ in Planck's original formulas, which gives values that differ by a factor of $\sqrt{2\pi}\simeq 2.51$.)

\begin{table}
	\centering
	\resizebox{\linewidth}{!}{
		\begin{tabular}{|l|c|c|c|c|c|c|c|c|c|}
			\toprule
			& $(\text{kilogram})^{-1}$ & $\sqrt{G}$ & $m_p^{-1}$ & $m_e^{-1}$ & $(\text{e volt})^{-1}$ & $(\text{kelvin})^{-1}$ & Cs cycle & meter & second \\
			\midrule
			$(\text{kilogram})^{-1}$ & \textbf{1.00} & 5.46E-08 & 1.67E-27 & \textbf{9.11E-31} & 1.78E-36 & 1.54E-40 & 6.78E-41 & 2.21E-42 & 7.37E-51 \\
			$\sqrt{G}$ & 1.83E+07 & \textbf{1.00} & \myhigh 3.07E-20 & \myhigh 1.67E-23 & 3.27E-29 & 2.82E-33 & \myhigh 1.24E-33 & \textbf{4.05E-35} & 1.35E-43 \\
			$m_p^{-1}$ & 5.98E+26 &\myhigh 3.26E+19 & \textbf{1.00} &\myhigh 5.45E-04 & 1.07E-09 & 9.18E-14 &\myhigh 4.05E-14 & 1.32E-15 & 4.41E-24 \\
			$m_e^{-1}$ & 1.10E+30 &\myhigh 5.99E+22 & \myhigh \textbf{1.84E+03} & \textbf{1.00} & 1.96E-06 & 1.69E-10 & \myhigh 7.44E-11 & 2.43E-12 & 8.09E-21 \\
			$(\text{e volt})^{-1}$ & 5.61E+35 & 3.06E+28 & 9.38E+08 & 5.11E+05 & \textbf{1.00} & 8.62E-05 & 3.80E-05 & 1.24E-06 & 4.14E-15 \\
			$(\text{kelvin})^{-1}$ & 6.51E+39 & 3.55E+32 & 1.09E+13 & 5.93E+09 & 1.16E+04 & \textbf{1.00} & 0.441 & 0.014 & 4.80E-11 \\
			Cs cycle & 1.48E+40 &\myhigh 8.05E+32 & \myhigh 2.47E+13 & \myhigh 1.34E+10 & 2.63E+04 & 2.267 & \textbf{1.00} & 0.033 & 1.09E-10 \\
			meter & 4.52E+41 & 2.47E+34 & 7.57E+14 & 4.12E+11 & 8.07E+05 & 69.503 & 30.663 & \textbf{1.00} & 3.34E-09 \\
			second & 1.36E+50 & 7.40E+42 & 2.27E+23 & 1.24E+20 & 2.42E+14 & 2.08E+10 & \textbf{9.19E+09} & \textbf{3.00E+08} & \textbf{1.00} \\
			\bottomrule
		\end{tabular}
	}
	\caption{\textbf{SI/Planck/Compton conversion factors and fundamental constants.} Where a $\sqrt{G}$ row or column intersects an SI unit, we have the Planck mass, temperature, time, etc., or its inverse. (The values here correspond to Planck's original definitions in terms of $h$, not $\hbar$.) Similarly, where $m_p$ and $m_e$ intersect SI units are the Compton wavenumber,  wavelength, period, etc.\ for the respective particles. The values of these numbers depend on the decisions of the BIPM. Ratios of natural scales, on the other hand,   are fixed by nature. Such ratios (highlighted) are, to within  geometric factors (e.g.\ $4\pi, \sqrt{2\pi}$), independent of human culture. }
\label{tab:NaturalConstants}
\end{table}

Interestingly, there is another, ``geometrized'' unit system that sets  $G$, $c$, and $k$  equal to 1 and  uses the centimeter as its reference scale \cite{misner_gravitation_1973}.  The geometrized system  emphasizes the fungibility between mass, energy, and length that is characteristic of general relativity.  (Conveniently, it also  humanizes astronomical masses; instead of $2\times 10^{33}$~g, the mass of the sun is $1.5\times 10^5$~cm $= 1.5$~km.) However, this particular emphasis necessarily de-emphasizes the fungibility between mass, energy, and frequency that is characteristic of quantum mechanics. In geometrized units, Planck's constant $h\simeq 1.64\times 10^{-65}$~cm$^2$, i.e.\ a scale with  the dimensions of area. The result is that time, distance, mass, energy, and temperature are all given in terms of the same unit, the centimeter.  As phenomena (e.g.\ diffraction) featuring fungibility between real- and reciprocal-space quantities are of subordinate concern with astronomical objects, this drawback does little apparent harm. The tension between unit systems where $G$ is taken to be a conversion factor and those where $h$ is  taken to be a conversion factor is a manifestation of the famous incompatibility between general relativity and quantum mechanics. That such seemingly contradictory systems can both be successfully employed highlights the arbitrariness and adaptability of unit systems in general.

Excepting the table concerning dimensionless quantities (Table~\ref{tab:dimensionless}), all of the tables shown in this article can be combined into one, ``grand'' table that relates these many units.  We have built this table up (but not displayed it, because of its size) according to one logical progression.  Now that we have a sense of its scope and structure, other sub-divisions  offer different perspectives. One illuminating sub-division is  according to geometric dimensionality: lengths, areas, volumes, etc..   When a table's scope is restricted in this way, the  constants that define it appear in their most recognizable form, since they are not raised to some  power. For instance, the newton, the watt, the hectare, and the field units are all areas or areal frequencies. We invite the reader to create this table, or some other table connecting the various units  commonly used in an area of interest.

The grand table embodies a constructive proof that all of the many units that it contains could be replaced by as few as one unit and powers thereof.  This  idea has a long history\cite{jackson_classical_1999}.  But summarizing the SI in this way also illuminates its meaning and significance.  From one point of view, the SI is a cross between a time capsule and a memorial: it archives more than 400 years of weighty scientific advances alongside some nearly arbitrary decisions,  while at the same time immortalizing almost 20 notable actors. However, if we instead look past its baroque facade and concentrate on the edifice's inner structure, we see an elegant distillation of the great theories of physics. The SI unites the classical units for time, length, mass, charge, temperature, energy, force, and field in a single, geometric framework. In so doing it manages, without any mathematics more complicated than arithmetic, to capture many of the foundational principles of electromagnetism,  thermodynamics, quantum mechanics, and relativity.

\bibliographystyle{naturemag19_etal-after-30}
\bibliography{lecture_notes_bibliography}

\begin{thebibliography}{10}
\expandafter\ifx\csname url\endcsname\relax
  \def\url#1{\texttt{#1}}\fi
\expandafter\ifx\csname urlprefix\endcsname\relax\def\urlprefix{URL }\fi
\providecommand{\bibinfo}[2]{#2}
\providecommand{\eprint}[2][]{\url{#2}}

\bibitem{inglis_si_2019}
\bibinfo{author}{Inglis, B.}, \bibinfo{author}{Ullrich, J.} \&
  \bibinfo{author}{Milton, M.}
\newblock \emph{\bibinfo{title}{{SI} {Brochure}: {The} {International} {System}
  of {Units} ({SI})}} (\bibinfo{publisher}{Bureau International des Poids et
  Mesures}, \bibinfo{year}{2019}), \bibinfo{edition}{9th} edn.
\newblock \urlprefix\url{https://doi.org/10.59161/AUEZ1291}.

\bibitem{wilczek_fundamental_2007}
\bibinfo{author}{Wilczek, F.}
\newblock \bibinfo{title}{Fundamental {Constants}} (\bibinfo{year}{2007}).
\newblock \urlprefix\url{http://arxiv.org/abs/0708.4361}.

\bibitem{duff_how_2015}
\bibinfo{author}{Duff, M.~J.}
\newblock \bibinfo{title}{How fundamental are fundamental constants?}
\newblock \emph{\bibinfo{journal}{Contemporary Physics}}
  \textbf{\bibinfo{volume}{56}}, \bibinfo{pages}{35--47}
  (\bibinfo{year}{2015}).
\newblock \urlprefix\url{https://doi.org/10.1080/00107514.2014.980093}.

\bibitem{1992Taylor}
\bibinfo{author}{Taylor, E.~F.} \& \bibinfo{author}{Wheeler, J.~A.}
\newblock \emph{\bibinfo{title}{Spacetime Physics : Introduction to Special
  Relativity}} (\bibinfo{publisher}{{W.H. Freeman}}, \bibinfo{address}{{New
  York}}, \bibinfo{year}{1992}).

\bibitem{navas_review_2024}
\bibinfo{author}{Navas, S.}, \bibinfo{author}{Amsler, C.},
  \bibinfo{author}{Gutsche, T.}, \bibinfo{author}{Hanhart, C.},
  \bibinfo{author}{Hernández-Rey, J.}, \bibinfo{author}{Lourenço, C.},
  \bibinfo{author}{Masoni, A.}, \bibinfo{author}{Mikhasenko, M.},
  \bibinfo{author}{Mitchell, R.}, \bibinfo{author}{Patrignani, C.},
  \bibinfo{author}{Schwanda, C.}, \bibinfo{author}{Spanier, S.},
  \bibinfo{author}{Venanzoni, G.}, \bibinfo{author}{Yuan, C.},
  \bibinfo{author}{Agashe, K.}, \bibinfo{author}{Aielli, G.},
  \bibinfo{author}{Allanach, B.}, \bibinfo{author}{Alvarez-Muñiz, J.},
  \bibinfo{author}{Antonelli, M.}, \bibinfo{author}{Aschenauer, E.},
  \bibinfo{author}{Asner, D.}, \bibinfo{author}{Assamagan, K.},
  \bibinfo{author}{Baer, H.}, \bibinfo{author}{Banerjee, S.},
  \bibinfo{author}{Barnett, R.}, \bibinfo{author}{Baudis, L.},
  \bibinfo{author}{Bauer, C.}, \bibinfo{author}{Beatty, J.},
  \bibinfo{author}{Beringer, J.}, \bibinfo{author}{Bettini, A.} \emph{et~al.}
\newblock \bibinfo{title}{Review of {Particle} {Physics}}.
\newblock \emph{\bibinfo{journal}{Physical Review D}}
  \textbf{\bibinfo{volume}{110}}, \bibinfo{pages}{030001}
  (\bibinfo{year}{2024}).
\newblock \urlprefix\url{https://link.aps.org/doi/10.1103/PhysRevD.110.030001}.

\bibitem{mohr_codata_2025}
\bibinfo{author}{Mohr, P.~J.}, \bibinfo{author}{Newell, D.~B.},
  \bibinfo{author}{Taylor, B.~N.} \& \bibinfo{author}{Tiesinga, E.}
\newblock \bibinfo{title}{{CODATA} recommended values of the fundamental
  physical constants: 2022}.
\newblock \emph{\bibinfo{journal}{Reviews of Modern Physics}}
  \textbf{\bibinfo{volume}{97}}, \bibinfo{pages}{025002}
  (\bibinfo{year}{2025}).
\newblock
  \urlprefix\url{https://link.aps.org/doi/10.1103/RevModPhys.97.025002}.

\bibitem{regan2020ahistoricalapproachelementaryphysics}
\bibinfo{author}{Regan, B.~C.}
\newblock \bibinfo{title}{An ahistorical approach to elementary physics}
  (\bibinfo{year}{2020}).
\newblock \urlprefix\url{https://arxiv.org/abs/2010.10271}.

\bibitem{lynch_history_1985}
\bibinfo{author}{Lynch, A.}
\newblock \bibinfo{title}{History of the electrical units and early standards}.
\newblock \emph{\bibinfo{journal}{IEE Proceedings A (Physical Science,
  Measurement and Instrumentation, Management and Education, Reviews)}}
  \textbf{\bibinfo{volume}{132}}, \bibinfo{pages}{564--573}
  (\bibinfo{year}{1985}).
\newblock
  \urlprefix\url{https://digital-library.theiet.org/doi/abs/10.1049/ip-a-1.1985.0097}.

\bibitem{misner_gravitation_1973}
\bibinfo{author}{Misner, C.~W.}, \bibinfo{author}{Thorne, K.~S.} \&
  \bibinfo{author}{Wheeler, J.~A.}
\newblock \emph{\bibinfo{title}{Gravitation}} (\bibinfo{publisher}{W.H. Freeman
  and Company}, \bibinfo{address}{New York}, \bibinfo{year}{1973}).

\bibitem{jackson_classical_1999}
\bibinfo{author}{Jackson, J.~D.}
\newblock \emph{\bibinfo{title}{Classical electrodynamics}}
  (\bibinfo{publisher}{Wiley}, \bibinfo{address}{New York},
  \bibinfo{year}{1999}), \bibinfo{edition}{3rd} edn.

\end{thebibliography}

\end{document}